\documentclass[aps,twocolumn,superscriptaddress,fleqn,showpacs,prl]{revtex4}
\usepackage{amsmath,amssymb}
\usepackage{graphicx}
\usepackage{bm}

\begin{document}

\title{Collective dynamical response of coupled oscillators with any network structure}
\author{Hiroshi Kori}
\email[corresponding author: ]{kori.hiroshi@ocha.ac.jp}
\affiliation{Division of Advanced Sciences, Ochadai Academic Production,
Ochanomizu Univeristy, Tokyo 112-8610, Japan}
\affiliation{PRESTO, Japan Science and Technology Agency, Kawaguchi
332-0012, Japan}
\author{Yoji Kawamura}
\affiliation{Earth Simulator Center, Japan Agency for Marine-Earth Science and Technology, Yokohama 236-0001, Japan}
\author{Hiroya Nakao}
\affiliation{Department of Physics, Graduate School of Sciences, Kyoto University, Kyoto 606-8502, Japan}
\author{Kensuke Arai}
\affiliation{Department of Physics, Graduate School of Sciences, Kyoto University, Kyoto 606-8502, Japan}
\author{Yoshiki Kuramoto}
\affiliation{Research Institute for Mathematical Sciences, Kyoto
University, Kyoto 606-8502, Japan}
\affiliation{Institute for Integrated Cell-Material Sciences, Kyoto University, Kyoto 606-8501, Japan}
\date{\today}
\begin{abstract}
 We formulate a reduction theory that describes the response of an
 oscillator network as a whole to external forcing applied nonuniformly
 to its constituent oscillators.  The phase description of multiple
 oscillator networks coupled weakly is also developed. General formulae
 for the collective phase sensitivity and the effective phase coupling
 between the oscillator networks are found. Our theory is applicable to
 a wide variety of oscillator networks undergoing frequency
 synchronization. Any network structure can systematically be treated.
 A few examples are given to illustrate our theory.
\end{abstract}


\maketitle

An assembly of coupled limit-cycle oscillators often behaves like a
single large oscillator.  This general scenario recurs in a wide variety
of rhythmic phenomena in living organisms, ranging from circadian
oscillations, cardiac rhythms to pathological phenomena such as epilepsy
and Parkinsonian disease
\cite{winfree80,kuramoto84,tass99,pikovsky01,reppert02}. Recent
experiments using electrochemical oscillators simulate such naturally
arising populations of oscillators in an idealized form
\cite{kiss02}.

Many previous studies have been devoted to answering how and under what
conditions oscillators mutually synchronize. In comparison, little
attention has been paid to investigating the dynamical response of an
oscillator network to external stimuli.  Such an inquiry would shed
light on mechanisms underlying biological functions (such as the phase
response curves of circadian rhythms \cite{johnson99}), external
control, and inter-network synchronization of oscillator networks
\cite{okuda91}. Establishing
a description of the collective dynamics on the basis of ``microscopic''
knowledge, i.e., the nature of the constituent oscillators and their
mutual coupling, presents a challenging task from a theoretical point of
view.  In particular, it would be ideal if the collective dynamics could
be described in terms of a single, suitably defined collective phase
$\Theta$ in a closed form.  A quantity of central importance will then
be {\em the phase sensitivity} of a population as a whole to external
weak stimuli \footnote{The phase sensitivity is the function of the
phase of an oscillator describing the phase shifts (compared to
unperturbed one) normalized by the intensity of weak pulsative
perturbation given to the oscillator
\cite{winfree67,winfree80,kuramoto84}. It is also sometimes called the
infinitesimal phase response curve \cite{izhikevich07}.}. This function has already been
derived \cite{kawamura08} for a system consisiting of a large assembly
of identical phase oscillators with global coupling, with each
oscillator being driven independently by random time-dependent noise.

In this Letter, we argue that there is yet another general situation
for which a similar phase description can be formulated.  In contrast to
Ref.~\onlinecite{kawamura08}, we consider noise-free but nonidentical
oscillators undergoing full frequency synchronization (in which all the
oscillators have exactly the same frequency due to coupling).
A similar situation has also been studied by Ko and Ermentrout for two
symmetrically coupled and globally coupled oscillators very recently
\cite{ko08}.  A distinct advantage of our present approach is that it
may deal with any system size, any connectivity, any heterogeneous
coupling, and nonuniform external forcing.  Moreover, the theory is
extended to include multiple populations simply by reinterpreting the
external stimuli applied to a given population as the coupling forces
originating from the other populations, which enables us to predict the
synchronization behavior between oscillator networks.  General formulae
for the collective phase sensitivity and the effective phase coupling
between the oscillator networks are found.

Consider a network of $N$ coupled limit-cycle oscillators under external
forcing. The forcing is generally nonuniform, i.e., individual oscillators receive
different inputs.  As is well known \cite{kuramoto84}, if the
heterogeneity of oscillators, the coupling between oscillators, and
external forcing are weak, the system is describable by the phase
equation
\begin{equation}
  \dot \phi_i = \omega_i + \sum_{j=1}^N \Gamma_{ij}( \phi_i - \phi_j ) +
  \epsilon Z(\phi_i) \xi_i(t). \label{phase_model}
\end{equation}
Here $\phi_i$ is the phase of the $i$th
oscillator ($i=1,\ldots,N$), $\omega_i$ its natural frequency,
and $\Gamma_{ij}$ the coupling force from the $j$th oscillator to
the $i$th oscillator.  The terms $\xi_i(t)$ and $Z(\phi_i)$ respectively
represent the time-dependent external force and the phase sensitivity of
the oscillator $i$ to external perturbation \footnote{When the forcing
is given by a vector $\bm \xi_i (t)$ in the original dynamical system
model before reduction, the corresponding forcing term in the
phase-reduced equation is generally given, unlike
Eq.(\ref{phase_model}), by a scalar product between this forcing vector
and the phase sensitivity vector $\bm Z(\phi_i)$ of the $i$th
oscillator \cite{kuramoto84}.  Here, $\bm Z(\phi_i)$ is the left
eigenvector with the zero eigenvalue defined for the linearized system
of the $i$th oscillator about its limit-cycle orbit.  However, to avoid
unnecessary complications, we have used a simpler form of the forcing
term in Eq.~(\ref{phase_model}) assuming that the original forcing
vector has only one nonvanishing component $\xi_i$.}.  Parameter
$\epsilon$ is the characteristic intensity of the external forcing.

Our aim is to establish the collective phase description for
Eq.~(\ref{phase_model}), i.e. to derive the dynamical equation for
a suitably defined macroscopic variable that describes the response to
external forcing. This is generally formulated under two basic
assumptions. (i) In the absence of external forcing a stable periodic
solution corresponding to a fully frequency-synchronized state exists,
and thus, the oscillator network behaves as a single large limit-cycle
oscillator. (ii) The external force is even weaker than the coupling
force, i.e. $\epsilon \ll 1$, so that the synchronized state is almost
unaltered under external forcing. Under these assumptions, the phase
reduction method \cite{kuramoto84} applicable to a weakly perturbed
oscillator can be applied (once again) to the oscillator network by
interpreting the unperturbed system as a single limit-cycle oscillator.


For convenience, we begin by rewriting Eq.~(\ref{phase_model}) in terms of the
$N$-dimensional state vector $\bm X = ( \phi_1, \phi_2, \ldots, \phi_N)$ as
\begin{equation}
  \dot{\bm X} = {\bm F}( {\bm X} ) + \epsilon {\bm p}( {\bm X}, t ), \label{X-eq}
\end{equation}
where $F_i(\bm X) = \omega_i + \sum_{j=1}^N \Gamma_{ij}( \phi_i - \phi_j
)$ and $p_i(\bm X, t) = Z(\phi_i) \xi_i(t)$.  A frequency synchronized
state (or, more exactly, a phase locked state) for $\epsilon=0$ is
found as a solution of $F_i(\bm X)=\Omega$ for all $i$.  This
``limit-cycle'' solution is denoted by
\begin{equation}
  \bm X_0(\Theta) = ( \Theta + \phi_1^0, \Theta + \phi_2^0, \ldots,
   \Theta + \phi_N^0 ), \ \Theta = \Omega t,
  \label{limit_cycle}
\end{equation}
where $\phi_i^0$ is constant.  We then extend the definition of $\Theta$
outside the limit-cycle orbit as a scalar field $\Theta(\bm X)$.  The identity $\dot
\Theta = (d\Theta/d\bm X)\cdot (d\bm X/dt)$ implies
\begin{equation}
 \dot \Theta = \frac{d\Theta}{d\bm X} \cdot \left\{\bm F(\bm X) + \epsilon \bm p(\bm
  X, t) \right\}.
 \label{Theta_full}
\end{equation}
As is usually done in the phase reduction process, we now adopt the
definition of $\Theta(\bm X)$ such that in the
absence of the forcing, $\dot\Theta=\Omega$ is identically satisfied,
i.e., $(d\Theta/d\bm X) \cdot \bm F(\bm X) = \Omega$.  We call
$\Theta(\bm X)$ {\em the collective phase} \footnote{ In our particular
system, $\Theta$ can be explicitly given for $\bm X \simeq \bm X_0$ as
$\Theta = \bm U^* \bm X + \Theta_0$ with some constant $\Theta_0$,
because $\dot \Theta=\bm U^* \dot{\bm X} = \bm U^* \bm F (\bm X_0) + \bm
U^* L \bm Y = \sum_{i=1}^N U_i^* \Omega=\Omega$. Correspondingly, one
may immediately obtain Eq.~(\ref{dot_Theta}) by applying $\bm U^*$ to
the lowest order description of Eq.~(\ref{X-eq}), $\dot {\bm X}=\bm
F(\bm X_0) + L\bm Y + \epsilon \bm p(\bm X_0,t)$, as well as by
following the normal phase reduction process.}.  Due to the assumption
of weak forcing, $(d\Theta/d\bm X) \cdot \bm p(\bm X,t)$ may be evaluated at
$\bm X=\bm X_0(\Theta)$. Then, to the lowest order in $\epsilon$
Eq. (\ref{Theta_full}) becomes
\begin{equation}
 \dot \Theta = \Omega + \epsilon \frac{d\Theta}{d\bm X_0}\cdot  \bm p(\bm
  X_0, t).
  \label{Theta_lowest}
\end{equation}
Let the linearized equation of $(\ref{X-eq})$ with $\epsilon=0$ be
$\dot{\bm Y} = L \bm Y$, where $\bm Y$ is the deviation defined by $\bm
Y=\bm X-\bm X_0$. Due to the symmetry of $\bm F(\bm X)$, the Jacobian
$L$ is a constant matrix, one eigenvalue of which equals zero with the
corresponding eigenvector given by $\bm U = d \bm X_0(\Theta)/ d \Theta
= (1,1,\ldots,1)$. We define row vector $\bm U^*$ as the left
zero-eigenvector of $L$, i.e., $\bm U^* L=0$, with the normalization
condition $\bm U^* \bm U$=1, or, $\sum_{i=1}^N U_i^\ast=1$.  It can then
be argued that $d\Theta/d\bm X_0$ becomes identical to $\bm U^*$
\cite{kuramoto84}.  Thus, Eq.~(\ref{Theta_lowest}) takes the form
\begin{equation}
 \dot \Theta =    \Omega+ \epsilon \bm \zeta(\Theta)\cdot \bm \xi(t),
  \label{dot_Theta}
\end{equation}
where $\bm \xi=(\xi_1,\xi_2,\ldots,\xi_N)$ and $\bm \zeta(\Theta)$,
which is interpreted as {\em the collective phase sensitivity}, is a
vector with components
\begin{equation}
 \zeta_i(\Theta) = U_i^* Z(\Theta+\phi_i^0).
  \label{cprc1}
\end{equation}
Equations (\ref{dot_Theta}) and (\ref{cprc1}) have clarified the
response of the collective mode of the synchronized network. There, the
forcing to the oscillator $i$ has turned out to be weighted by
$U_i^*$. We thus call $\bm U^*$ {\em the weight vector} henceforth.

In what follows, we often consider uniform external forcing, $\xi_i(t) = \xi(t)$ for all $i$.
In such a case, Eq.~(\ref{dot_Theta}) is further reduced to $\dot \Theta
= \Omega + \epsilon \zeta(\Theta) \xi(t)$, where $\zeta(\Theta)$ is the
collective phase sensitivity defined for uniform forcing, $\zeta( \Theta
) = \sum_{i=1}^N U_i^* Z(\Theta+\phi_i^0)$. Generally speaking, $\zeta(\Theta)$
deviates from $Z(\phi)$ more significantly as the phases of the constituent
oscillators are more widely distributed. 

An analytic formula of weight vector $\bm U^*$ is given as follows.
The elements of $L$ are given by
\begin{equation}
  L_{ij} = \delta_{ij} \sum_{k \ne i}^N \Gamma_{ik}' \left( \phi_i^0 - \phi_k^0 \right) 
  - ( 1 - \delta_{ij}) \Gamma_{ij}'\left( \phi_i^0 - \phi_j^0 \right).
\end{equation}
Note that the relation $L \bm U = \sum_{j=1}^N L_{ij} = 0$ holds.  We define the
$(i,i)$-cofactor of $L$ as $M_i = {\rm det}\, L(i,i)$, which is the
determinant of the submatrix $L(i,i)$, that is $L$ with the $i$-th row and
column removed. One can prove that, for any $L$ with $\sum_{j=1}^N
L_{ij} = 0$, the row vector $(M_1, M_2, ..., M_N)$ is a left
zero-eigenvector of $L$, i.e., $\sum_{j=1}^N M_j L_{ji} = 0$
\cite{biggs97}.  Using the normalization condition, we obtain the
algebraic expression of $\bm U^*$ as
\begin{equation}
 \bm U^*=(M_1, M_2, \ldots, M_N)/M,
  \label{U}
\end{equation}
where $M=\sum_{i=1}^N M_i$. This expression is valid for any network
structure. Note that, for networks with both bidirectional connection
patterns and symmetric coupling functions, $L$ is symmetric and 
$U_i^*$ is trivially $1/N$ for all $i$, which is the case even for
networks with strongly heterogeneous connectivity including scale-free networks.
In many cases, however, $L$ is asymmetric and $\bm U^*$ is
heterogeneous.


Next, we formulate the collective phase description of multiple networks
of phase oscillators.  We are concerned with the case in which external
forcing is absent, while $\epsilon \xi_i(t)$ in the last term in
Eq.~(\ref{phase_model}) is interpreted as representing the coupling
force coming from oscillators of other networks.  For clarity, we consider
a simple system in which two identical networks composed of $N$
oscillators, called group A and group B, are uniformly coupled (the
extension to a more general case, e.g., weakly heterogeneous multiple
groups, is straightforward). The dynamical equations of the system are given by
\begin{equation}
 \begin{aligned}
  \dot \phi_i^{\rm A} &= \omega_i + \sum_{j=1}^N \Gamma_{ij}( \phi_i^{\rm A} - \phi_j^{\rm A} )
  + \epsilon Z(\phi_i^{\rm A}) \xi(\{\phi^{\rm B}_k\}), \\
  \dot \phi_i^{\rm B} &= \omega_i + \sum_{j=1}^N \Gamma_{ij}( \phi_i^{\rm B} - \phi_j^{\rm B})
  + \epsilon Z(\phi_i^{\rm B}) \xi(\{\phi^{\rm A}_k\}),
\end{aligned}
 \label{coupled_phase_models}
\end{equation}
where $\phi_i^{\rm X}$ is the phase of the oscillator $i$ in group X
(X=A, B), and $\xi(\{\phi^{\rm X}_k\})$ denotes a function of
$\phi_1^{\rm X},\phi_2^{\rm X}, \ldots, \phi_N^{\rm X}$, which
represents the uniform coupling force coming from group X.  Denoting the
collective phases of the respective groups by $\Theta_{\rm A}$ and
$\Theta_{\rm B}$, we obtain the resulting phase equation in the form
\begin{equation}
\begin{aligned}
 \dot \Theta_{\rm A} &= \Omega + \epsilon \zeta(\Theta_{\rm A})
   \xi(\{\Theta_{\rm B}+\phi_k^0\}), \\
 \dot \Theta_{\rm B} &= \Omega + \epsilon \zeta(\Theta_{\rm B})
   \xi(\{\Theta_{\rm A}+\phi_k^0\}). \\
\end{aligned}  
 \label{AB-coupling}
\end{equation}
Noting that the dominant time-dependence of $\Theta_{\rm A,B}$ is $\Omega t$,
we may obtain the effective coupling $\gamma( \Theta_{\rm A} - \Theta_{\rm B} )$
between the groups by time-averaging of Eq.~(\ref{AB-coupling}) over the
common period $2\pi/\Omega$.
Putting $\Theta_{\rm A,B} = \Omega t +
\theta_{\rm A,B}$, we perform the averaging as
\begin{align}
 \gamma &( \Theta_{\rm A} - \Theta_{\rm B} ) = \gamma( \theta_{\rm A} - \theta_{\rm B} ) = \notag \\
&\frac{1}{2\pi} \int_0^{2\pi} d(\Omega t)
   \zeta(\Omega t +\theta_{\rm A}) \xi(\{\Omega t +\theta_{\rm B}+\phi_k^0\}),
   \label{collective_phase_coupling}
\end{align}
where $\gamma$ rather than $\Gamma$ has been used to
indicate that this coupling function acts between the groups.
In this way, we succeeded in deriving the collective phase equation
in the simple form
\begin{equation}
\begin{aligned}
   \dot \Theta_{\rm A} &= \Omega + \epsilon \gamma( \Theta_{\rm A} -
 \Theta_{\rm B} ),\\
   \dot \Theta_{\rm B} &= \Omega + \epsilon \gamma( \Theta_{\rm B} - \Theta_{\rm A} ).
\end{aligned}
   \label{collective_phase_model}
\end{equation}

We give two examples to illustrate our theory.

\begin{figure}
\includegraphics[width=5.5cm]{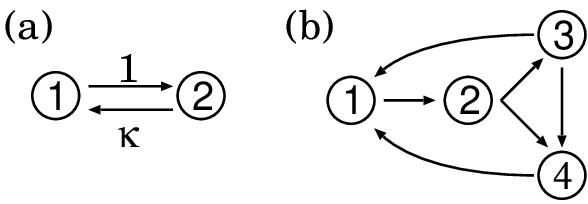} 
\caption{Examples of weight vector
 $\bm U^*$. Using Eq.~(\ref{U}), we find (a) $U^*_1:U^*_2 = 1:\kappa$
 and (b) $U^*_1:U^*_2:U^*_3:U^*_4 = 2:4:3:1$.}  \label{fig:network}
\end{figure}

{\em Example I: the dependence of weight vector $\bm U^*$ on the
connectivity}. We consider a network of identical oscillators with
homogeneous coupling, i.e., $\omega_i=\omega$ and $\Gamma_{ij}(\Delta
\phi)=A_{ij} \Gamma(\Delta \phi)$ undergoing perfect phase
synchrony. Here, $A$ is the adjacency matrix describing the
connectivity, which is generally asymmetric and weighted.  In such a
system, the weight vector depends solely on the network architecture.
By rescaling $t$ and $\epsilon$, we put $\Gamma'(0)=-1$ without loss of
generality. In perfect phase synchrony, i.e., $\phi_i^0=\phi^0$ for all
$i$, $L$ is a Laplacian matrix generalized for asymmetric and weighted
networks, given by $L_{ij}=-A_{ij}$ for $i \ne j$ and $L_{ii}=\sum_{j
\ne i}^N A_{ij}$. We consider two small networks in which the
weight vector $\bm U^*$ is easily calculated via the algebraic
expression, Eq.~(\ref{U}). Figure \ref{fig:network}(a) is a weighted
network, where the weight vector is found to be a simple reflection of
the connection weights. Figure \ref{fig:network}(b) is a non-weighted
network but its adjacency matrix is asymmetric.  It is worth noticing
that oscillator 2 is more influential than oscillator 3, although they
have locally the same topological properties: one inward and two outward
connections. In general, the weight vector depends on the global
topology.

{\em Example II: collective phase sensitivity and group synchronization
in limit-cycle oscillators}.  We illustrate that the collective phase
sensitivity of a group of coupled oscillators varies with intra-group
coupling strength.  We then consider two groups of coupled oscillators
with an additional inter-group coupling of fixed strength, and show that
a nontrivial qualitative change in the synchronization behavior between
the groups occurs when the individual collective phase sensitivities
changes as a result of modifications to the intra-group coupling
strength.

\begin{figure}
\includegraphics[width=8cm]{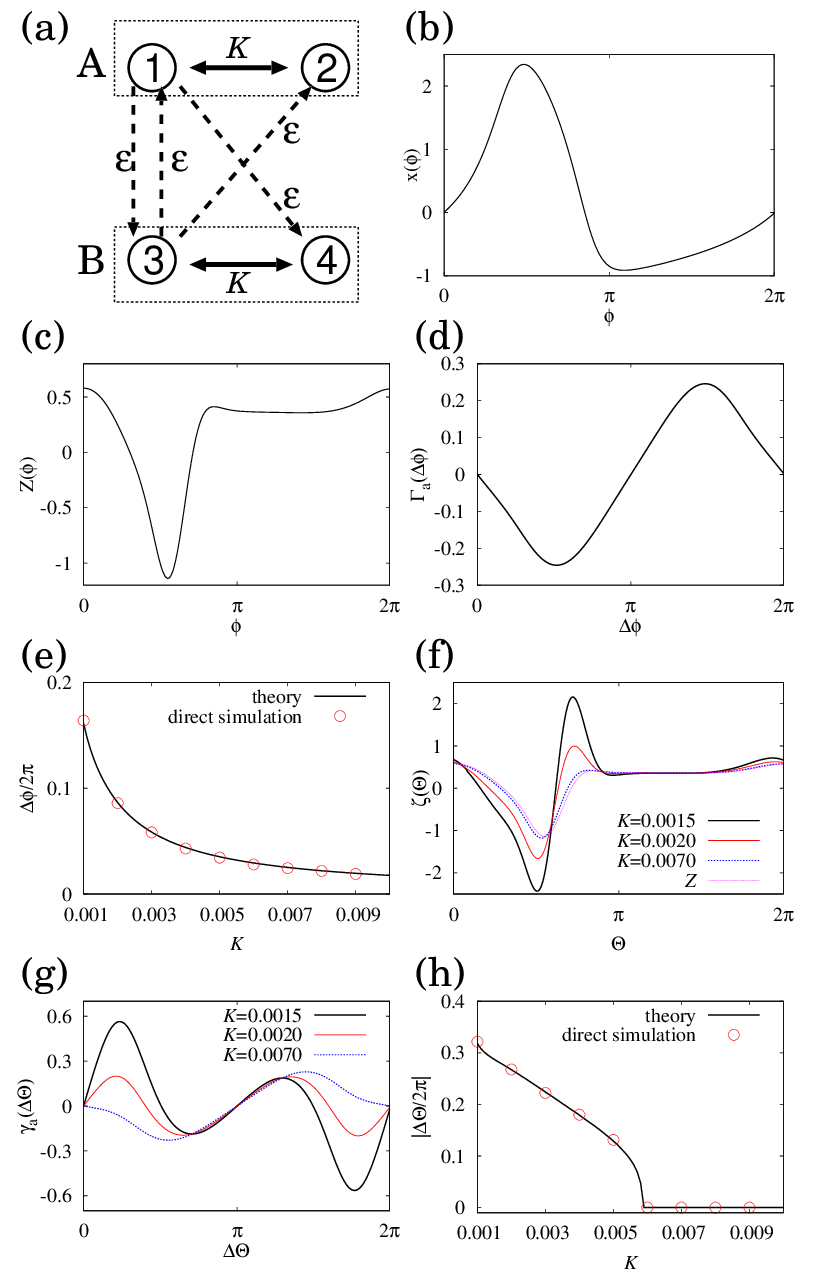} 
\caption{(color online) Results for a network of
 limit-cycle oscillators. (a) Network structure under
 consideration. (b) Waveform $x(\phi)$ and (c) phase sensitivity
 $Z(\phi)$ of an individual oscillator. (d) Antisymmetric part of the
 coupling function, $\Gamma_{\rm a}(\Delta \phi)$. (e) Phase difference $\Delta
 \phi^0$ between oscillators in each group. (f) Collective phase
 sensitivity $\zeta(\Theta)$ defined for uniform forcing. For
 $K=0.007,0.002,0.0015$, $(U_1^*,U_2^*)$ is respectively about
 $(1.35,-0.35),(2.87,-1.87)$ and $(2.89,-1.89)$.  (g) Antisymmetric part
 of the collective coupling function, $\gamma_{\rm a}(\Delta \Theta)$. (h) Phase
 difference $\Delta \Theta$ between the groups. } \label{fig:hr}
\end{figure}

As schematically illustrated in Fig.~\ref{fig:hr}(a), we consider the
system in which a pair of identical groups A and B, each of which
consists of two coupled limit-cycle oscillators, are mutually
coupled. We use the Hindmarsh-Rose model as the limit-cycle oscillator,
a model originally proposed as a neural model. The system reads
\begin{equation}
 \dot x_i = 3x_i^2-x_i^3+y_i-\mu_i + \sum_{j=1}^4 D_{ij} x_j, \,\, \dot y_i = 1-5x_i^2-y_i.
  \label{hr}
\end{equation}
Here, the coupling matrix $D$ is given as $D_{12,21,34,43}=K,
D_{11,22,33,44}=-K, D_{13,23,31,41}=\epsilon$ with $K$ and $\epsilon$
being the coupling intensities for intra- and inter-groups,
respectively. We assume $1 \gg K \gg \epsilon=1.0 \times 10^{-5}$. We
set $\mu_1=\mu_3=-3.000$ and $\mu_2=\mu_4=-3.001$, corresponding to
$\omega_{1,3} \simeq 1.80476, \omega_{2,4} \simeq 1.80443$, and thus
$\Delta \omega \simeq 3.3 \times 10^{-4}$.  The wave form $x(\phi)$ and the
phase sensitivity $Z(\phi)$ of an isolated oscillator obtained
numerically are shown in Figs. \ref{fig:hr}(b) and \ref{fig:hr}(c),
respectively.

The corresponding phase model is given by
Eq.~(\ref{coupled_phase_models}), where
$\Gamma_{ij}(\Delta \phi)=K \Gamma(\Delta \phi)$, $\xi(\{\phi_k^{\rm
A}\})=x(\phi_1)$, and $\xi(\{\phi_k^{\rm B}\})=x(\phi_3)$.
From the phase reduction theory, the coupling
function is calculated as 
$\Gamma(\phi_1-\phi_2)=\frac{1}{2\pi}\int_{0}^{2\pi} d(\omega t)
Z(\phi_1+\omega t)\{x(\phi_2+\omega t)-x(\phi_1 + \omega t)\}$.  For
convenience, we display $\Gamma_{\rm a}(\Delta \phi) \equiv
\Gamma(\Delta \phi) - \Gamma(-\Delta \phi)$ in
Fig. \ref{fig:hr}(d). The phase difference $\Delta \phi^0$ between the
oscillators $1$ and $2$ of a synchronized state is found as a stable
solution of $\dot \phi_1=\dot \phi_2$ (where $\epsilon=0$ is assumed),
and thus, a solution of $\Gamma_{\rm a} (\Delta \phi^0)=\Delta
\omega/K$. The predicted phase difference is plotted in
Fig. \ref{fig:hr}(e) as a curve. It agrees well with numerical data
obtained through direct numerical integration of Eq.~(\ref{hr}). Using
$Z(\phi)$, $\Delta \phi^0$, $\Gamma(\Delta \phi)$, and its derivative
$\Gamma'(\Delta \phi)$ obtained numerically, we can calculate $U_1^*,
U_2^*$, and $\zeta(\Theta)$. The results are shown in
Fig. \ref{fig:hr}(f) and its caption. For large $K$ (compared to $\Delta
\omega$), $\zeta(\Theta)$ is indistinguishable from $Z(\phi)$. As $K$
decreases, $\Delta \phi^0$ becomes larger, resulting in considerable
variation in $\zeta(\Theta)$.

Given $\zeta(\Theta)$, the synchronization behavior between groups is
now predicted. The collective coupling function $\gamma(\Delta \Theta)$ is
calculated from Eq.~(\ref{collective_phase_coupling}) where
$\xi(\phi)=x(\phi)$ in the system under consideration. For convenience,
we display the antisymmetric part of $\gamma(\Delta \Theta)$ in
Fig.~\ref{fig:hr}(g). Putting $\dot \Theta_{\rm A}=\dot \Theta_{\rm B}$,
we find the stable phase locking solution between groups, $\Delta \Theta
\equiv \Theta_{\rm A}- \Theta_{\rm B}$. Predicted $\Delta
\Theta$ is exhibited by the curve in Fig. \ref{fig:hr}(h), implying that
the in-phase solution becomes unstable at around $K=0.006$ (via a
pitch-fork bifurcation) and the out-of-phase solution appears
below. Phase difference $\Delta \Theta$ (or equivalently,
$\phi_1-\phi_3$) obtained from direct numerical integration of
Eq.~(\ref{hr}) is plotted in Fig. \ref{fig:hr}(h), which convinces us
of the precision of the phase description.

In summary, we have formulated the response of the collective
phase to weak external forcing given to constituent oscillators and
the phase description for interacting oscillator networks. The
present theory is valid for a wide class of weakly coupled oscillator
networks undergoing full frequency synchronization, and thus, a broad applicability
would be expected.



\end{document}